\RequirePackage{lineno}
\pdfoutput=1%For ARXIV
\UseRawInputEncoding
\documentclass[aps,prb,preprint,groupedaddress,showkeys]{revtex4}
\usepackage[pdftex]{graphicx}
\usepackage[colorlinks=true,linkcolor=blue,citecolor=blue]{hyperref}
\graphicspath{{figures/}}
\DeclareGraphicsExtensions{.pdf,.jpeg,.png}

\usepackage{url}

\begin{document}

\def\papertitle{3D EAGAN: 3D edge-aware attention generative adversarial network for prostate segmentation in transrectal ultrasound images}
\title{\papertitle}

\author{Mengqing Liu}
\affiliation{School of Information Engineering, Nanchang Hangkong University, Nanchang, Jiangxi, China.}
\author{Xiao Shao}
\affiliation{School of Computer Science, Nanjing University of Information Science and Technology, Nanjing, Jiangsu, China.}
\author{Liping Jiang}
\affiliation{The First Affiliated Hospital of Nanchang University, Nanchang University, Nanchang, Jiangxi, China.}
\author{Kaizhi Wu}
\affiliation{School of Information Engineering, Nanchang Hangkong University, Nanchang, Jiangxi, China.
Key Laboratory of Jiangxi Province for Image Processing and Pattern Recognition, Nanchang Hangkong University, Nanchang, China.}

%\date{\today}

\begin{abstract}	% <= 300 words
\textbf{Background:} Segment prostates from transrectal ultrasound (TRUS) images plays an essential role in the diagnosis and treatment of prostate cancer. However, traditional segmentation methods are time-consuming and laborious.
To address this issue, there is an urgent need to develop computer algorithms that can automatically segment prostates from TRUS images, which makes it become the direction and form of future development.

\textbf{Purpose:} Automatic prostate segmentation in TRUS images has always been a challenging problem, since prostates in TRUS images have ambiguous boundaries and inhomogeneous intensity distribution. Although many prostate segmentation methods have been proposed, they still need to be improved due to the lack of sensibility to edge information. Consequently, the objective of this study is to devise a highly effective prostate segmentation method that overcomes these limitations and achieves accurate segmentation of prostates in TRUS images.

\textbf{Methods:} A 3D edge-aware attention generative adversarial network (3D EAGAN)-based prostate segmentation method is proposed in this paper, which consists of an edge-aware segmentation network (EASNet) that performs the prostate segmentation and a discriminator network that distinguishes predicted prostates from real prostates. The proposed EASNet is composed of an encoder-decoder-based U-Net backbone network, a detail compensation module, four 3D spatial and channel attention modules, an edge enhance module, and a global feature extractor. The detail compensation module is proposed to compensate for the loss of detailed information caused by the down-sampling process of the encoder. The features of the detail compensation module are selectively enhanced by the 3D spatial and channel attention module. Furthermore, an edge enhance module is proposed to guide shallow layers in the EASNet to focus on contour and edge information in prostates. Finally, features from shallow layers and hierarchical features from the decoder module are fused through the global feature extractor to predict the segmentation prostates.

\textbf{Results:} The proposed method is evaluated on the TRUS image dataset consists of 56 patients, which achieves the mean Dice of 92.80±0.75\%, Jaccard of 87.01±0.42\%, HD of 4.64±0.69mm, Precision of 93.11±0.62\%, and Recall of 92.42±1.00\%, respectively. Experiments have been conducted on our TRUS image dataset to compare with seven state-of-the-art segmentation methods. Specifically, our proposed method outperforms other methods on all metrics, including Dice, Jaccard, HD, Precision, and Recall.

\textbf{Conclusion:} A novel 3D edge-aware attention generative adversarial network-based prostate segmentation method is proposed. The proposed method consists of an edge-aware segmentation network and a discriminator network. Experimental results demonstrate that the proposed method has achieved satisfactory results on 3D TRUS image segmentation for prostates.
\end{abstract}

\maketitle

\section{Introduction}
Prostate cancer is one of the most common cancers diagnosed in men \cite{[1]}. Since early-stage prostate cancer can be effectively controlled, early detection and interventions are crucial to the diagnosis and treatment planning of prostate diseases. Transrectal ultrasound (TRUS) images have the characteristics of good real-time performance and cost-effectiveness, which makes them widely used in the diagnosis and treatment of prostate cancer. Hence, accurate segmentation of prostates from TRUS images plays a vital role in the fields of prevention, diagnosis, and treatment of prostate cancer. Traditionally, TRUS images were manually segmented by experienced doctors, and it is a time-consuming and laborious process that requires the expert experience of doctors. Hence, developing computer algorithms to realize automatic prostates segmentation from TRUS images holds significant importance.

Generally, traditional prostate segmentation methods \cite{[2],[3],[4],[5],[6],[7],[8],[9],[10],[11],[12],[13],[14]} utilize hand-crafted features (e.g., shape statistics) for segmentation, but these hand-crafted features are low-level semantic features and are not capable to capture complex features from real prostates. Recently, benefiting from the feature extraction ability of deep convolutional neural networks (DCNN), many semantic segmentation methods have emerged, which aim to assign a set of predefined classes to each pixel in images. Long et al. \cite{[15]} proposed the fully convolutional network (FCN)-based method for image segmentation tasks, which is an end-to-end architecture to automatically classify images to different classes. Ronneberger et al. \cite{[16]} proposed an encoder-decoder-based U-Net architecture for medical semantic segmentation, which utilizes the skip connection to integrate low-level features extracted by the encoder into the decoder. Inspired by these novel architectures, a large number of DCNN-based prostate segmentation methods \cite{[17],[18],[19],[20]} were proposed.

Although these methods achieved great improvements over traditional methods, further improvements are still lacking. Different from other semantic segmentation tasks (e.g., indoor scenes and street scenes), TRUS images have weak boundaries, low signal-to-noise ratio, and large differences in contrast and resolution. Specifically, TRUS images have ambiguous boundaries caused by poor contrast between the prostate and surrounding tissues. Hence, current methods which directly adopt semantic segmentation models (e.g., FCN and U-Net) to segment prostates would lack the sensitivity to ambiguous boundaries and inhomogeneous intensity distribution of prostates. Therefore, it is quite challenging to accurately segment the boundary of prostates. 

In this paper, a novel 3D edge-aware attention generative adversarial network (3D EAGAN)-based prostate segmentation method is proposed. The proposed method consists of an edge-aware segmentation network (EASNet) and a discriminator network. The EASNet aims to produce prostate segmentation results and the discriminator is designed to distinguish the predicted prostates from the ground-truth prostates. The EASNet is composed of an encoder-decoder-based U-Net backbone network, a detail compensation module, four 3D spatial and channel attention modules (3D SCAM), an edge enhance module, and a global feature extractor. Since the down-sampling of the encoder in EASNet would cause information loss, the detail compensation module is proposed to introduce rich detail contextual information to the encoder, which is pre-trained on a large-scale medical data set 3DSeg-8 \cite{[med3d]} to learn rich details and texture information. Due to the detail compensation module would contain some irrelevant features with prostates, the 3D SCAM is proposed to selectively utilize the features that can reflect more prostate details from the channel and spatial dimensions. To further assist the EASNet to generate more accurate prostate margins, an edge enhance module is proposed to guide shallow layers in the EASNet to focus on contour and edge information in prostates. Finally, the enhanced low-level features from the encoder and hierarchical features from the decoder are fused to global feature extractor to obtain the final segmentation results. In summary, this paper has the following main contributions:

\begin{itemize}
\item[1. ]  A novel framework 3D EAGAN for improving prostate segmentation is proposed, which adopts a detail compensation module to learn rich detail information of prostates and an edge enhance module to guide the network focus on edge information of prostates.

\item[2. ] Since prostates in TRUS images have ambiguous boundaries, an edge enhance module is introduced to further guide shallow layers in the encoder to focus on the prostate edges without adding extra computation cast during the inference process.

\item[3. ] A 3D spatial and channel attention module is proposed to adaptively enhance the features that can reflect prostate details by considering interdependencies among channel and spatial dimensions.
\end{itemize}

The remainder of this paper is organized as follows. Section II reviews the automatic prostate segmentation methods in TRUS images, including traditional prostate segmentation methods and learning-based methods. Section III presents the details of the proposed 3D EAGAN, including the overall network architecture, detail compensation module, 3D spatial and channel attention module, edge enhance module, global feature extractor, and loss functions. Section IV presents the experimental results of the proposed method, including the single dataset experiments and visualization of feature maps. Section V presents the conclusion of this study.

\section{Related work}
In this section,  prostate segmentation methods in TRUS images are reviewed, including traditional prostate segmentation methods and deep learning-based methods.

\subsection{Traditional prostate segmentation methods}
According to the way of feature extraction, the segmentation method can be divided into hand-crafted-based prostate segmentation methods and deep learning-based ones. Traditional hand-crafted-based prostate segmentation methods utilize carefully designed hand-crafted features to detect the shape and edge of prostates. The shape statistics belong to the mainstream of traditional segmentation methods. Ladak et al. \cite{[2]} proposed a semi-automatic segmentation method based on 2D ultrasound images, which first utilized shape statistics to detect prostates. To detect the edge of prostates, Pathak et al. \cite{[3]} developed an edge detection algorithm to depict the prostate edges. Shen et al. \cite{[4]} employed Gabor filter sets to characterize prostate boundaries and reconstructed Gabor features to guide deformable segmentation. Yan et al. \cite{[5]} learned the shape statistical information of the local domain to segment prostates. Santiago et al. \cite{[6]} employed an active shape model (ASM) to improve the robustness in the presence of outliers. Although these methods have achieved more promising segmentation performance than traditional manual segmentation methods, these methods are performed on 2D TRUS images, which would lack the correlation between different image slices and 3D image context. 

To effectively enhance the correlation between different TRUS image slices and 3D image context, many 3D prostate segmentation methods are proposed. Ghanei et al. \cite{[7]} proposed a 3D deformable surface model to segment ultrasound images. Wang et al. \cite{[8]} proposed two semi-automatic segmentation methods by using 2D ultrasound images to achieve 3D prostate. Hu et al. \cite{[9]} employed a semi-automatic segmentation by using an efficient deformable mesh. Gong et al. \cite{[10]} used deformable models for the automatic segmentation of prostates. Qiu et al. \cite{[11]} proposed a novel globally optimized method to segment 3D prostate images. Previously methods utilize shape information of prostates to enhance the segmentation performance, but the shape of prostates varies greatly, which would lose the specificity of individual cases and lead to a decrease in prediction accuracy. Different from these shape statistics-based methods, many other prostate segmentation methods treat the segmentation task as a classification task. Ghose et al. \cite{[12]} applied the principal component analysis and random forest classification in machine learning to implement prostate segmentation. Zhan et al. \cite{[13]} proposed a deformable model for automatic prostate segmentation by shape and texture statistics. To augment training samples, Yang et al. \cite{[14]} proposed a 3D TRUS image segmentation method for the prostate based on a patch-based feature learning framework. Although these hand-crafted-based methods have achieved promising prediction accuracy, the hand-crafted features are shallow and not capable to obtain high-level semantic information in images, resulting in the lack of prostate boundary information.

\subsection{Deep learning-based prostate segmentation methods}
Recently, deep learning technology has achieved great success in various image processing tasks, including image classification \cite{[36]}, image enhancement \cite{[37]}, and semantic segmentation \cite{[38]}. Benefiting from features automatically learned by convolutional neural networks, many deep learning-based prostate segmentation methods have been proposed. Ghavami et al. \cite{[17]} employed a U-Net-based method for automatic prostate segmentation, which replace the convolutional layers in the original U-Net with residual network unit blocks to enhance the feature representation ability. To solve the problem of information loss in traditional shape models, Yang et al. \cite{[18]} used the recurrent neural network to learn the shape prior information of prostates. Wang et al. \cite{[19]} proposed a 3D deep neural network-based prostate segmentation method, which first utilized the 3D feature pyramid network (FPN) to extract multi-level features. Then, an attention mechanism was proposed to adaptive fuse different features and pay attention to the prostate region. Lei et al. \cite{[20]} employed the V-Net-based backbone network to extract primary features, and the 3D supervision mechanism was integrated into the network training to speed up the network convergence. Pellicer-Valero et al. \cite{[43]} proposed a Densenet-resnet-based 3D prostate segmentation method. In addition, to improve the robustness of the network, some techniques, e.g., deep supervision, checkpoint ensembling, and neural resolution enhancement, are also integrated into the network training process. 

Currently, generative adversarial networks (GAN) have made impressive progress in many computer vision tasks \cite{[44],[45]}. Generally, GAN is composed of two parts: a generator network $G$ and a discriminator network $D$. $G$ aims to generate more real samples and $D$ is designed to distinguish the real samples from the generated samples by $G$. The training stage of $G$ and $D$ is to optimize the minimax game by using the objective function,
\begin{equation}
\mathop{min} \limits_G \mathop{max} \limits_D F(D, G)=E_{x\sim P_{data}}[log(D(x))]+E_{y\sim P_{y}}[log(1-D(G(y)))],
\end{equation}
where $G$ denotes the generator network; $D$ denotes the discriminator network; $P_{y}$ represents the distribution of the noise; $P_{data}$ represents the distribution of real data. To further enhance the prostates segmentation performance, methods \cite{[46],[47]} adopted the GAN for the prostates segmentation task. Dong et al. \cite{[46]} utilized the adversarial training strategy for the prostate segmentation task. The generator is composed of a set of U-Nets and the discriminator is the fully convolutional network. Wang et al. \cite{[47]} employed the GAN to automatically segment prostates, which consists of a Densenet-based generator and a multi-scale discriminator.

\section{Method}

Due to ambiguous boundaries and inhomogeneous intensity distribution of prostates in TRUS images, segmenting prostates from TRUS images is still a challenging task. To effectively segment prostates from TRUS images, a 3D EAGAN-based prostate segmentation method is proposed in this paper. The architecture of the proposed 3D EAGAN is shown in FIG. 1, it consists of two parts: an edge-aware segmentation network and a discriminator network. The edge-aware segmentation network aims to segment more accurate prostates to fool the discriminator, and the discriminator network is expected to distinguish predicted prostates from real prostates. Then, the edge-aware segmentation network and the discriminator network are described in detail.

\begin{figure}[htbp]
  \centering
  \includegraphics[width=6.7in]{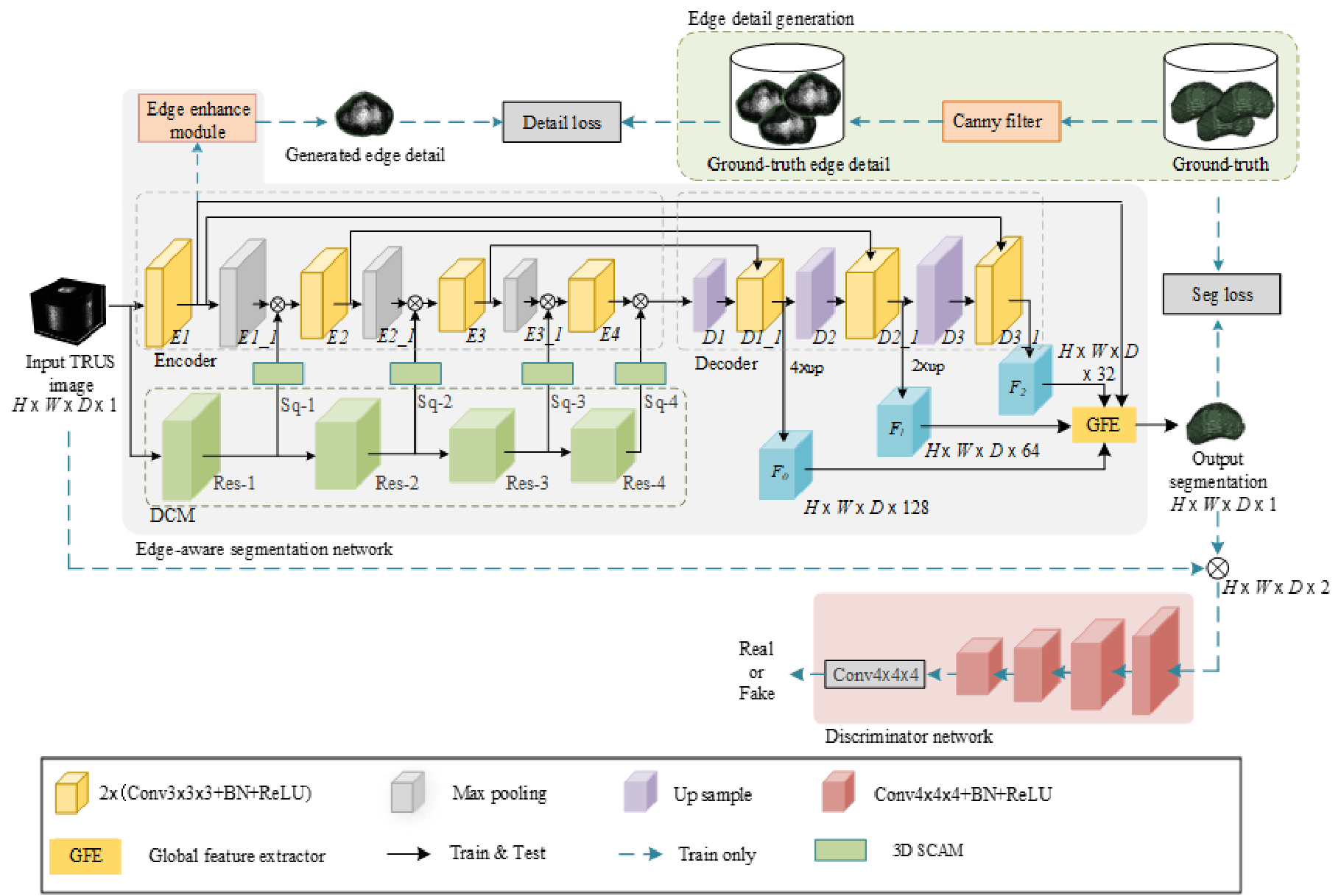}
  \caption{The framework of the proposed method.}
\end{figure}

\subsection{Edge-aware segmentation network}

The network architecture of the proposed edge-aware segmentation network is shown in FIG. 1, which is composed of an encoder-decoder-based U-Net, a detail compensation module (DCM), four 3D spatial and channel attention modules (3D SCAM), an edge enhance module, and a global feature extractor (GFE). For the input TRUS images, the encoder-decoder-based U-Net is first used to extract primary features from input. To compensate for the loss of detail information during the U-Net forward propagation, the input TRUS images are also fed to the DCM to introduce rich local detail information to the encoder module. To reduce the computational complexity, 1$\times$1$\times$1 convolutional layers Sq-$i$ ($i\in${1, 2, 3, 4}) are adopted to the output of the DCM to squeeze the channel dimension. Then, the extracted detail information is refined by the 3D SCAM, and feature maps from the encoder module and 3D SCAM at each resolution stage are fused through concatenation operation. In addition, an edge enhance module is utilized to guide the shallow layers in the encoder to focus on contour and edge information. Finally, the GFE is used to integrate hierarchical features from the decoder and shallow layers from the encoder to generate more realistic prostates.

\subsection{Encoder-decoder-based U-Net}

Since the U-Net structure \cite{[16]} has shown strong feature representation ability in the image segmentation task, the encoder-decoder-based U-Net structure is adopted as the backbone network. The encoder module aims to extract low-level and high-level semantic features from TRUS images, while the decoder module is designed to progressively combine contextual information from different levels to generate the output segmentation.

More specifically, the encoder module is composed of four Conv-BN-ReLU (CBR) modules and three max-pooling layers. Each CBR module consists of two groups of 3$\times$3$\times$3 convolutional layers followed by a batch normalization (BN) layer and a ReLU activation. Max pooling layers are applied to gradually down-sample the resolutions of feature maps to half to reduce the computational complexity and improve the inference speed of the network. The decoder module consists of three CBR modules and three up-sampling layers. The up-sampling layers are used to gradually recover the resolutions of feature maps to match the resolutions of the input image.

\subsection{Detail compensation module}
Due to the requirement of fast inference and low resource consumption, the U-Net structure utilizes the down-sampling operation in the encoder to progressively decrease the resolutions of the input images. However, the down-sampling operation would cause the loss of detailed contextual information. Since prostates in TRUS images have ambiguous boundaries, the loss of detailed contextual information would inevitably cause the degradation of segmentation ability.

Hence, the transfer learning-based detail compensation module is proposed, and it is built on a ResNet-34 \cite{[36]} pre-trained on the large-scale medical dataset 3DSeg-8 \cite{[med3d]}. The network-based transfer learning technique aims to solve the problem of limited training data \cite{[48]}. Its underlying assumption is that the internal layers of a convolutional neural network (CNN) are not specific to a particular task, e.g., the shallow layers from an image classification CNN are sensitive to the detail information (e.g., edge and texture features). Specifically, the network-based transfer learning technique usually pre-trains a network on the source task $T_{s}$, and features learned from the source task are transferred to the target task $T_{t}$ to enhance the robustness of the network. The architecture of the DCM is shown in FIG. 1. First, the ResNet-34 is trained on the 3DSeg-8 to learn the features of different organs. Then, the internal layers Res-$i$ ($i\in${1, 2, 3, 4}) of the pre-trained ResNet-34 are transferred to the prostate segmentation task to learn more abundant detailed contextual information.

\subsection{3D spatial and channel attention module}
Recently, attention mechanisms have been widely used in many computer vision tasks, which can effectively boost the performance of DCNNs. In medical segmentation tasks, many methods have applied attention mechanisms to make the network focus on target regions. Since the features extracted from DCM contain rich detailed information, but also include some non-prostate features. To further enhance and refine more important features, attention mechanisms are introduced to adaptive filter out non-prostate features and focus on important features by exploring the relationship of features between the spatial and channel dimensions. The architecture of the proposed 3D SCAM is shown in FIG. 2, and it is composed of a spatial attention module and a channel attention module. 

\begin{figure}[htbp]
  \centering
  \includegraphics[width=5.0in]{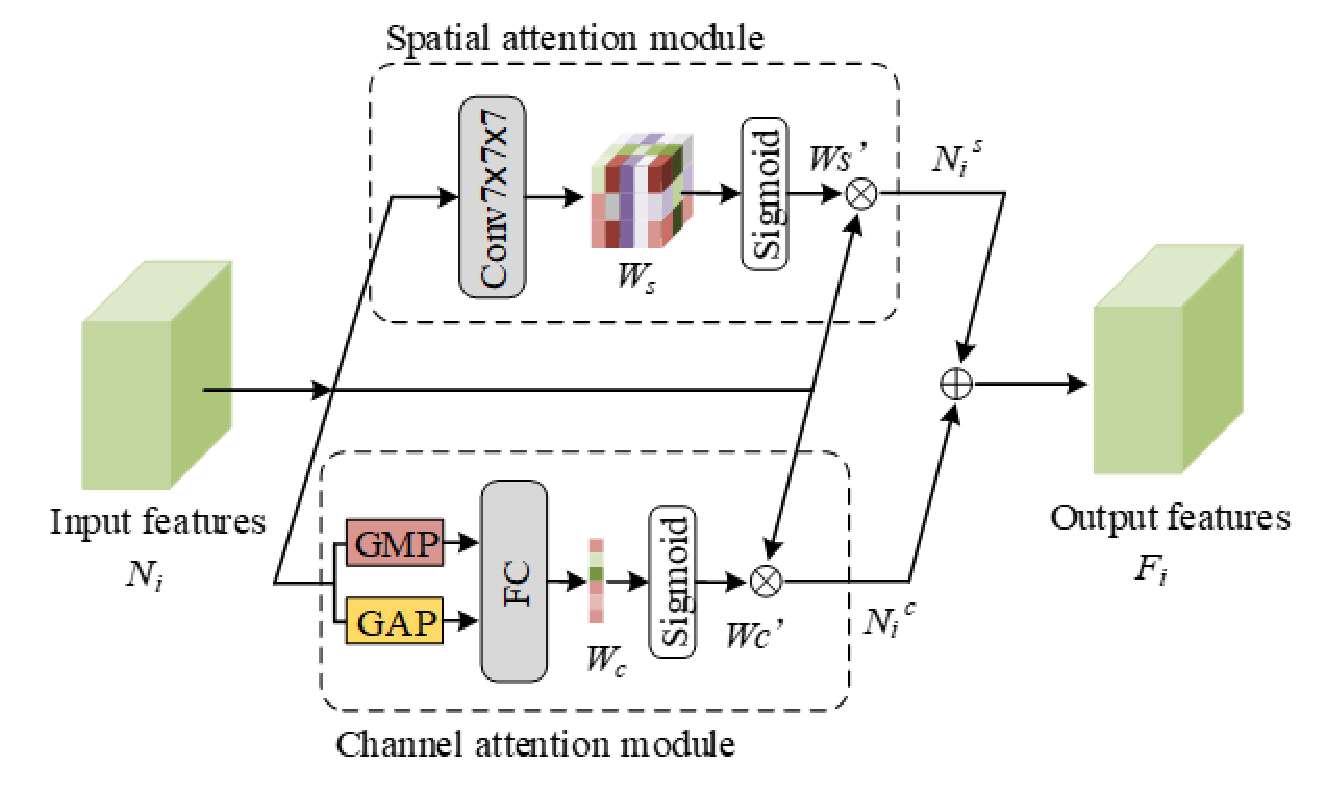}
  \caption{The network structure of the 3D SCAM.}
\end{figure}

Specifically, the feature maps $N_{i}$ from Res-$i$ ($i\in${1, 2, 3, 4}) in DCM are fed into the 3D SCAM. Given the input feature maps $N_{i}\in R^{H\times W\times D\times C}$ ($i\in${1, 2, 3, 4}), they are first fed into the spatial attention module and the channel attention module to generate the weight scores, respectively. The spatial attention module consists of a convolutional layer with a kernel size of 7$\times$7$\times$7 and a Sigmoid layer. The 7$\times$7$\times$7 convolutional layer is used to calculate the spatial weight scores $W_{s}$ from input feature maps $N_{i}$. Then, the Sigmoid layer is adapted to constrain the weight scores $W_{s}$ to be between [0, 1] to obtain $W_{s}^{'}$. Finally, the calculated spatial weight scores $W_{s}^{'}$ are multiplied by the input feature maps $N_{i}$ to obtain the adaptive feature maps $N_{i}^{s}$. The working mechanism of the spatial attention module is shown as follows,
\begin{equation}
\left\{
\begin{array}{l}
W_{s}=Conv_{1}(N_{i}) ,\\
W_{s}^{'}=\alpha (W_{s}) ,\\
N_{i}^{s}=W_{s}^{'}\times N_{i},\\
\end{array} \right.
\end{equation}
where $Conv_{1}()$ denotes the convolutional layer with kernel size 7$\times$7$\times$7. $\alpha$() denotes the Sigmoid layer. $\times$ denotes the element-wise multiply. For the channel attention module, it consists of a global max pooling (GMP) layer, a global average pooling (GAP) layer, a fully connected (FC) layer, and a Sigmoid layer. Given the input feature maps $N_{i}$, the GMP and GAP layers are first used to squeeze the features maps to the $N_{i}^{’}\in R^{H\times W\times D\times 1}$ along the channel dimension. Then, the feature maps $N_{i}^{'}$ are fed into the FC layer to calculate the channel weight scores $W_{c}$. The Sigmoid layer is adapted to constrain the $W_{c}$ to be between [0, 1] to obtain $W_{c}^{'}$. Finally, the input feature maps $N_{i}$ are calculated with the channel weight scores $W_{c}^{'}$ to obtain the adaptive feature maps $N_{i}^{c}$. The working mechanism of the channel attention module is shown as follows,
\begin{equation}
\left\{
\begin{array}{l}
W_{c}=GMP(N_{i})\otimes GAP(N_{i}) ,\\
W_{c}^{'}=\alpha (W_{c}) ,\\
N_{i}^{s}=W_{c}^{'}\times N_{i},\\
\end{array} \right.
\end{equation}
where $GMP()$ and $GAP()$ denote the global max pooling layer and global average pooling layer, respectively. $\otimes$ denotes the concatenation operation. $\alpha()$ denotes the Sigmoid layer. $\times$ denotes the element-wise multiply. Finally, feature maps $N_{i}^{s}$ and $N_{i}^{c}$ are fused through the element-wise addition operation. This process can be described as,
\begin{equation}
F_{i}=N_{i}^{s}+N_{i}^{c} ,
\end{equation}
where $+$ denotes the element-wise addition operation.

\subsection{Edge generation guidance of low-level features}
Since prostates in TRUS images have ambiguous structure boundaries, current methods fail to accurately predict the structure boundary of prostates. To enhance the sensitivity to edge details, an edge enhance module is proposed to guide shallow layers in the encoder module to focus on the edge details of prostates.

\begin{figure}[htbp]
  \centering
  \includegraphics[width=6.5in]{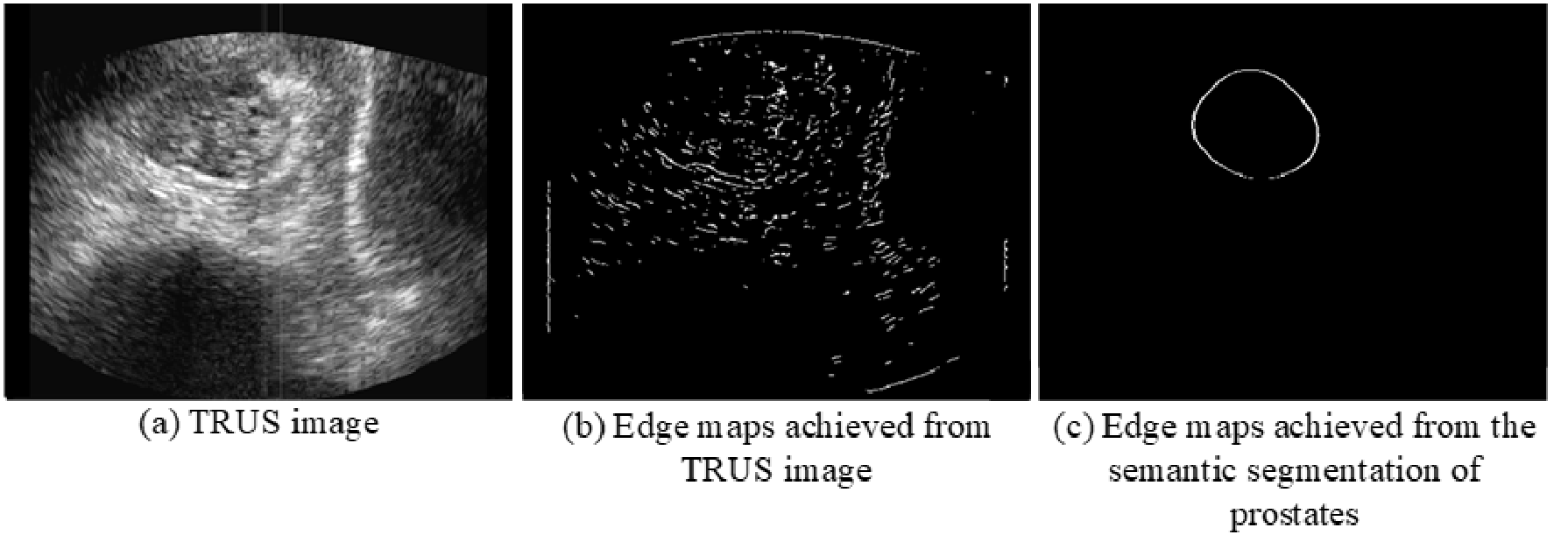}
  \caption{Visual comparisons of different ways to obtain edge maps.}
\end{figure}

To accurately obtain the ground-truth edge maps of prostates, two ways are tested to obtain the edge maps by using the Canny algorithm \cite{[49]}. First, edge maps are obtained from TRUS images. Second, edge maps are obtained from the ground-truth semantic segmentation prostate images. To intuitively show the difference between the edge maps calculated from TRUS images and the ground-truth semantic segmentation prostate images, the visualization of the different edge maps is shown in FIG. 3. It can be observed that edge maps directly achieved from TRUS images contain useless information. On the contrary, edge maps obtained from the ground-truth semantic segmentation prostate images can accurately reflect the edge of prostates. Hence, the ground-truth edge maps are generated from the ground-truth semantic segmentation prostate images by using the Canny algorithm.

With the generated ground-truth edge maps, the edge enhance module is used to guide the low-level layers in the encoder module to focus on learning the prostate boundary. The edge enhance module is composed of a 3$\times$3$\times$3 convolutional layer for feature extraction and a 1$\times$1$\times$1 convolutional layer to reduce the channel dimension. Finally, the learned edge features are fused with the hierarchical features in the decoder module for the final prediction.

\begin{figure}[htbp]
  \centering
  \includegraphics[width=6.5in]{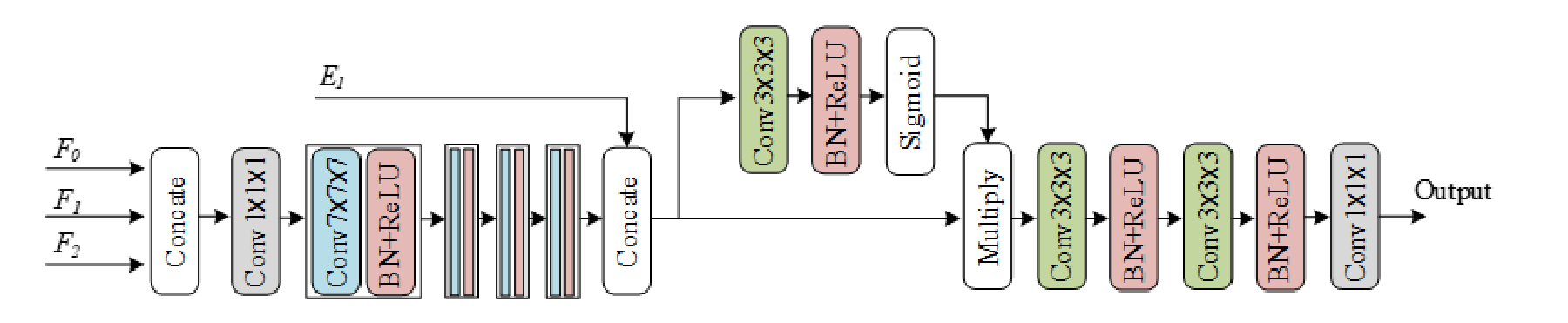}
  \caption{The network architecture of the global feature extractor.}
\end{figure}

\subsection{Global feature extractor}
To obtain more accurate segmentation performance and achieve a precise prostate edge, a global feature extractor is proposed. The architecture of the proposed global feature extractor is shown in FIG. 4. Specifically, multi-layer features $F_{i}$ ($i\in${0, 1, 2}) are first fused through the concatenation operation. The 1$\times$1$\times$1 convolutional layers are adopted to decrease the number of the channel dimension, which aims to reduce the computational complexity. Then, four 7$\times$7$\times$7 convolutional blocks are designed to build a density connection between feature maps and per-pixel classifier, which enhances the capability to handle different shapes and sizes. Motivate by previous work \cite{[31]}, to enhance the sensitively to edge information, low-layer features $E_{1}$, which are enhanced by the edge enhance module, are also introduced to the global feature extractor. To further help the network select more important features, a spatial attention structure is adopted, which is similar to the spatial attention module in the 3D SCAM. In the spatial attention structure, the 3$\times$3$\times$3 convolutional block is utilized for the feature extraction and the Sigmoid operation is used to constrain the value of the weight scores to be between [0, 1]. The weight scores calculated by the spatial attention structure are multiplied by the fused features to obtain the selective features. Finally, two 3$\times$3$\times$3 convolutional blocks are used for the final feature extraction and the 1$\times$1$\times$1 convolutional layer aims to map the channel dimension to match the channel of predicted prostates.

\subsection{The discrimination network}

The discriminator of the traditional GAN utilizes the whole image as input to conduct the discrimination, which only outputs one value to determine whether the generated image is real or false. Different from mapping the whole image to one value, PatchGAN \cite{[50]} extracts features from input image, and then maps input image into $N\times N$ matrix by the full convolution structure. Benefiting from PatchGAN structure, it can effectively enhance the attention to each area of the image. Hence, to achieve a better discriminative effect, PatchGAN is adopted as the discriminator network of the proposed 3D EAGAN.

\subsection{The loss function}
In deep learning tasks, the loss function plays a vital role in the neural network model training process. An elaborately designed loss function can effectively speed up the convergence of the model training and improve the prediction accuracy of the model. In the training process, the edge-aware segmentation network $G$ and the discriminator network $D$ are optimized by the minimax game. The objective function for training the edge-aware segmentation network is defined as:
\begin{equation}
minL(D)=[D(x,y)-l_{r}]^{2}+[D(x,G(x))-l_{f}]^{2} ,
\end{equation}
where $G()$ denotes the edge-aware segmentation network and $D()$ denotes the discriminator network. $x$ and $y$ denote input TRUS images and ground-truth labels, respectively. $l_{r}$ and $l_{f}$ represent the real label and fake matrix label with constant elements one and zero, respectively. The objective function for training the discriminator network is defined as:
\begin{equation}
maxL(G)=[D(x,G(x))-l_{t}]^{2}+ \alpha \cdot l_{dice}[y,G(x)]+\beta \cdot l_{dice}[y_{e},\hat{y_{e}}] ,
\end{equation}
where $l_{t}$ is the matrix with constant elements one. $y_{e}$ and $\hat{y_{e}}$ denote the predicted edge maps and ground-truth edge maps, respectively. $l_{dice}$ represents the Dice loss, which is widely used in medical image segmentation tasks. $\alpha$ and $\beta$ represent the hyper-parameters that control the impact of the loss function. According to extensive experiments, when $\alpha$ and $\beta$ are set to 1 and 0.5, respectively, the proposed method achieves the best prediction performance.

\section{Experiments}
In this section, the experimental setups are first introduced, including experimental environments and implementation tools, datasets, and evaluation metrics. Then, experiments are performed to compare the proposed method with other medical segmentation methods. Finally, the ablation study is conducted to verify the effectiveness of components in our method.

\subsection{Experimental setups}

\subsubsection{Experimental environments and implement tools}

The proposed method is programmed with Python 3.7 and implemented by PyTorch 1.2.0. To train the network, the training and testing process are performed on NVIDIA GeForce RTX 3090 GPU.

\subsubsection{Implementation details}

For the training stage, due to the limited GPU memory, input TRUS images are down-sampled with the size of 88$\times$112$\times$112. For the proposed 3D EAGAN, both the edge-aware segmentation network and the discriminator network are trained using the Adam optimizer \cite{[51]} with the parameters $\lambda_{1}$=0.9, $\lambda_{2}$=0.999, and the learning rate is initialized as 0.00001.

\subsubsection{Dataset}

The TRUS images are obtained through a mechanically assisted biopsy system used by collaborating radiologists [CR]\cite{[52]} of Western University. The study was reviewed and approved by a larger human subject research ethics of Western University.

The proposed method is evolution on a TRUS image dataset consisting of 56 patients. We acquired one 3D TRUS image from each patient. These 3D TRUS images are acquired with an end-firing 5-9 MHz TRUS transducer probe (Philips Medical Systems, Seattle, WA). The 3D TRUS image contains 350$\times$448$\times$448 voxels with a voxel size of 0.19$\times$0.18$\times$0.18$mm^{3}$. The data is processed using spatial and intensity distribution normalization.

\subsubsection{Compared methods}
To verify the effectiveness of the proposed method, seven state-of-the-art medical segmentation methods are used to conduct the experiments, including 3D FCN \cite{[15]}, 3D U-Net \cite{[16]}, Skip-Densenet \cite{[53]}, Deeplabv3+ \cite{[54]}, DAF 3D \cite{[19]}, Vox2Vox \cite{[55]}, and Chen et al. \cite{[56]}.

\subsubsection{Evaluation metrics}
As following previous work\cite{[17],[18],[19],[20],[43],[46],[47]}, five evaluation metrics are used to measure the segmentation performance of our proposed method, including Dice Similarity Coefficient (Dice), Jaccard Index (Jaccard), Hausdorff Distance (HD, in voxel), Precision, and Recall. 

The Dice is used to evaluate the similarity between predicted prostates and the ground truth ones,
\begin{equation}
Dice(P,G)=\frac{2|P|\cap |G|}{|P|+|G|} ,
\end{equation}
where $P$ and $G$ denote the predicted prostates and the ground-truth prostates. $|\cdot|$ represents the number of voxels. The value of Dice is in the range of [0, 1], the higher value denotes better segmentation performance.

The Jaccard is used to evaluate the similarity between predicted prostates and the ground truth ones,
\begin{equation}
Jaccard(P,G)=\frac{|P|\cap |G|}{|P|\cup|G|} ,
\end{equation}
where $P$ and $G$ denote the predicted prostates and the ground-truth prostates. $|\cdot|$ represents the number of voxels. The value of Dice is in the range of [0, 1], the higher value denotes better segmentation performance.

The HD is utilized to evaluate the distance between predicted prostates and the ground truth ones,
\begin{equation}
\left\{
\begin{array}{l}
h(A,B)= \mathop{max} \limits_{a\subset A}{\mathop{min} \limits_{b\subset B}||a-b||}, \\
h(B,A)= \mathop{max} \limits_{b\subset B}{\mathop{min} \limits_{a\subset A}||b-a||}, \\
HD(A,B)=max\{h(A,B),h(B,A)\}, \\
\end{array} \right.
\end{equation}
where $||\cdot||$ represents the distance paradigm between predicted prostates and the ground truth ones. The lower value of HD represents better segmentation performance.

The precision is used to evaluate the proportion of samples with a predicted value of one and a true value of one among all samples with a predicted value of one,
\begin{equation}
Precision(P,G)=\frac{Area(P\cap G)}{Area(P)} ,
\end{equation}
where $P$ and $G$ denote the predicted prostates and the ground-truth prostates. The value of precision is in the range of [0, 1], the higher value denotes better segmentation performance.

The Recall is used to evaluate the proportion of samples with a predicted value of one and a true value of one among all samples with a true value of one,
\begin{equation}
Recall(P,G)=\frac{Area(P\cap G)}{Area(G)} ,
\end{equation}
where $P$ and $G$ denote the predicted prostates and the ground-truth prostates. The value of Recall is in the range of [0, 1], the higher value denotes better segmentation performance.

\subsection{Comparison to state-of-the-art methods}
To verify the effectiveness of the proposed method, the segmentation performance of the proposed method is quantitatively evaluated through comparisons to seven state-of-the-art segmentation methods. The experimental results are shown in TABLE I. It can be observed that the proposed 3D EAGAN outperforms other methods on the used metrics. Specifically, 3D EAGAN achieves the mean Dice of 92.80\%, Jaccard of 87.01\%, HD of 4.64mm, Precision of 93.11\%, and Recall of 92.42\%, respectively. Compare with the traditional method of 3D FCN, our proposed method outperforms it by a large margin. The 3D FCN utilizes progressive down-sampling to reduce the resolution of the input image, which would lead to the loss of detailed information and degrades the segmentation performance. The 3D U-Net utilizes skip connection layers to effectively combine different features of the encoder and the decoder, which makes it achieve better segmentation performance than 3D FCN. The proposed DCM can effectively introduce abundant detailed information to the encoder, which compensates for the loss of detailed information caused by the down-sampling process. Hence, our proposed method improves the 3D U-Net by 6.46\%, 10.77\%, 6.13mm, 6.28\%, and 7.23\% on Dice, Jaccard, HD, Precision, and Recall, respectively. Compare with the DAF 3D, which also utilizes the attention mechanism to make the network focus on prostates, our proposed EAGAN improves it for 2.88\% on Dice, 5.26\% on Jaccard, 1.84mm on HD, 2.33\% on Precision, 2.30\% on Recall, respectively. The improvements benefit from the proposed 3D SCAM can not only focus on features in the spatial domain but also stress essential features in the channel domain. Compare with the GAN-based Vox2Vox, our proposed 3D EAGAN improves it by 2.48\% on Dice, 4.25\% on Jaccard, 2.59mm on HD, 2.07\% on Precision, and 1.08\% on Recall, respectively. Benefiting from the proposed edge enhance module, the proposed 3D EAGAN can pay attention to edge information, which leads to the improvement of segmentation performance.

\begin{table}[]
\caption{Quantitative results for prostate segmentation of different methods. The best results are marked in boldface.}
\centering
\begin{tabular}{p{80pt}p{80pt}p{80pt}p{80pt}p{70pt}p{70pt}}
\hline
Method  & Dice(\%) & Jaccard(\%) & HD(mm) & Precision(\%) & Recall(\%)   \\ \hline
   3D FCN \cite{[15]}  & 84.12$\pm$ 3.02    & 72.46$\pm$2.40  & 11.96$\pm$5.89 & 86.49$\pm$2.68 & 84.08$\pm$ 2.87   \\
   Chen et al. \cite{[56]} & 85.32$\pm$2.62	& 75.04$\pm$2.27	& 10.82$\pm$3.62	& 87.39$\pm$2.56	& 85.08$\pm$2.74 \\
   3D U-Net \cite{[16]} & 86.34$\pm$2.07 &76.24$\pm$2.64 &	10.77$\pm$4.05 &	86.83$\pm$3.22&	85.19$\pm$ 2.11 \\
   Skip-Densenet \cite{[53]} &
88.90$\pm$1.87&	80.56$\pm$2.19 &	8.95$\pm$2.63 &	90.25$\pm$1.94&	88.13$\pm$1.90 \\
Deeplabv3+ \cite{[54]} & 89.29$\pm$2.27&	80.96$\pm$2.53 &	6.89$\pm$1.82&	90.17$\pm$2.12&	88.39$\pm$ 1.75 \\
DAF 3D \cite{[19]} & 89.92$\pm$1.75 &	81.75$\pm$2.67 &	6.48$\pm$1.61 &	90.78$\pm$1.27&	90.12$\pm$1.58 \\
Vox2Vox \cite{[55]} & 90.32$\pm$1.57 &	82.76$\pm$1.99 &	7.23$\pm$2.20	& 91.04$\pm$1.17	& 91.34$\pm$1.46 \\

3D EAGAN	& \textbf{92.80$\pm$0.75}	& \textbf{87.01$\pm$0.42}	& \textbf{4.64$\pm$0.69}	& \textbf{93.11$\pm$0.62}&	\textbf{92.42$\pm$1.00} \\
        \hline
\end{tabular}
\end{table}

\begin{figure}[htbp]
  \centering
  \includegraphics[width=6.2in]{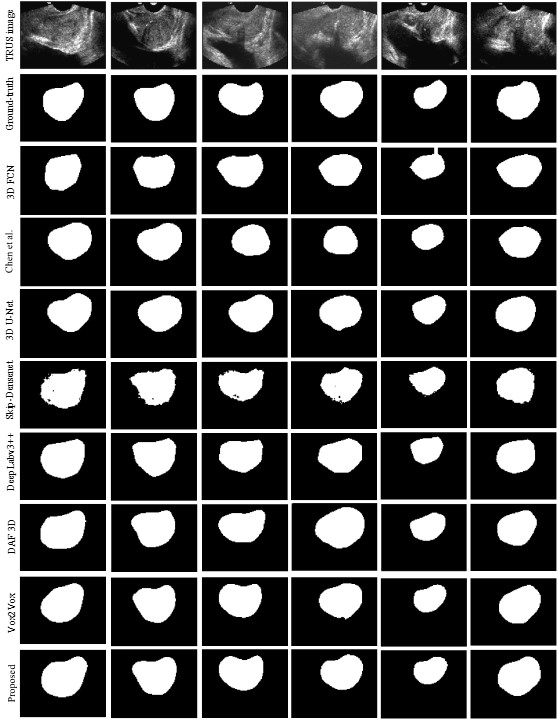}
  \caption{Visualization results of different methods on TRUS dataset.}
\end{figure}

To further verify the effectiveness of the proposed method, the 2D slice visualization results of prostate segmented by different methods are shown in FIG. 5.
The first row of images is the prostate TRUS images of different samples, the second row of images is the real prostate label image, and the rest of the images are the visualization of the segmentation effect of different methods.
For 3D FCN and Chen et al. methods, these methods use continuous downsampling to reduce the image resolution, which cause the loss of detail information. Hence, the results of these methods have a large difference between the predicted prostate and ground-truth one.
For Skip-Densenet and Deeplabv3++, there is also a large gap between the segmentation results and ground-truth ones, and there is a certain lack of segmentation results. 
The Vox2Vox method achieves better segmentation results than other methods, the reason is that it uses generative discriminator training to train the whole network.
Different from these methods, the prostate images segmented by the proposed method is closer to the ground-truth ones.
In summary, the reasons for satisfactory segmentation results of proposed method are: (1) The proposed edge enhance module can effectively enhance the perception of shallow features for prostate edge information, thereby improving the network's segmentation accuracy for prostates. 
(2) The proposed 3D SCAM enhances more important features in the spatial and channel dimensions through the attention mechanism, which makes the network pays more attention to the prostate region.

\subsection{Ablation study}

\subsubsection{Ablation study on the proposed detail compensation module} 
As discussed before, the detail compensation module is adopted in the edge-aware segmentation network to introduce abundant detail information to the network. The ablation study experiments are conducted to verify the effectiveness of the detail compensation module. The experimental results are shown in TABLE II. The proposed method without using the detail compensation module is denoted as “3D EAGAN w/o DCM”. On the contrary, “3D EAGAN w/ DCM” represents the detail compensation module used in the proposed method. It can be observed that with the use of the detail compensation module, the segmentation performance is significantly improved.

To further verify the effectiveness of the detail compensation module, feature maps are extracted from the encoder module in the edge-aware segmentation network. The visualization of the feature maps is shown in FIG. 6. It can be observed that with the use of the detail compensation module, abundant detail information can be introduced to the encoder module, which can enhance the robustness of the proposed method, and it consists with experimental results in TABLE II.

\begin{table}[]
\caption{Evaluation of using the detail compensation module.}
\centering
\begin{tabular}{p{115pt}p{70pt}p{70pt}p{70pt}p{70pt}p{70pt}}
\hline
Method  & Dice(\%) & Jaccard(\%) & HD(mm) & Precision(\%) & Recall(\%)   \\ \hline
   3D EAGAN w/o DCM  & 91.08$\pm$ 1.51    & 83.64$\pm$1.72  & 5.46$\pm$0.93 & 90.43$\pm$1.62 & 91.79$\pm$1.45   \\
3D EAGAN w/ DCM	& \textbf{92.80$\pm$0.75}	& \textbf{87.01$\pm$0.42}	& \textbf{4.64$\pm$0.69}	& \textbf{93.11$\pm$0.62}&	\textbf{92.42$\pm$1.00} \\
        \hline
\end{tabular}
\end{table}

\begin{figure}[htbp]
  \centering
  \includegraphics[width=6.5in]{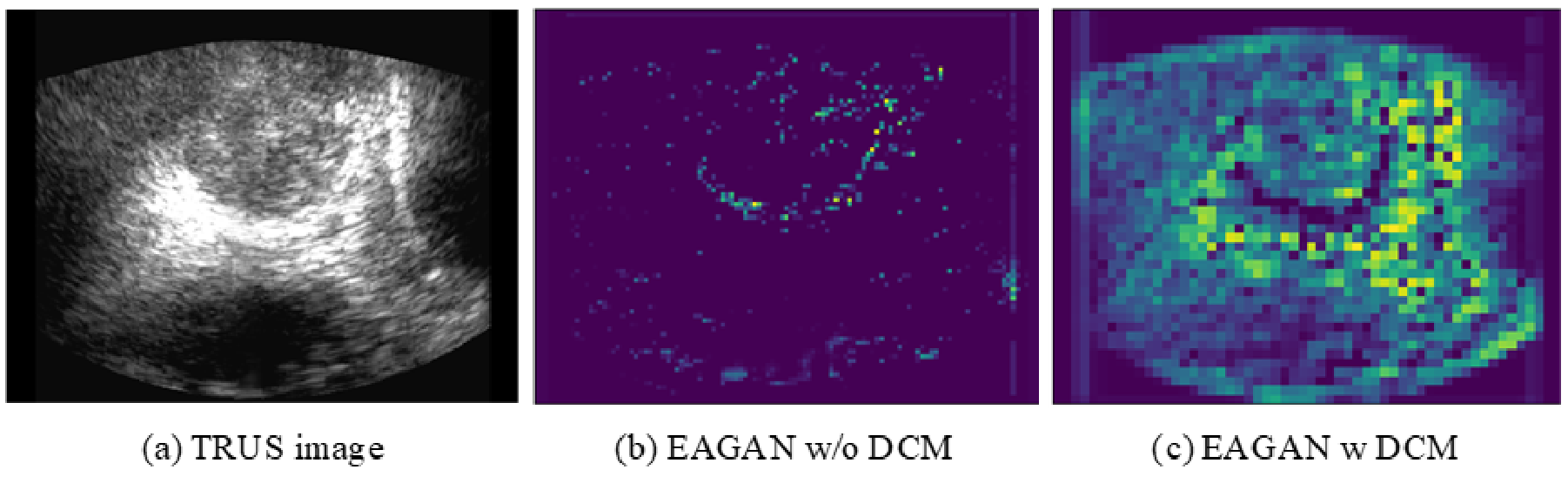}
  \caption{Visualization results of the feature maps extracted from the encoder module in the edge-aware segmentation network.}
\end{figure}

\subsubsection{Ablation study on the proposed 3D spatial and channel attention module}

The 3D spatial and channel attention module is added to the edge-aware segmentation network to selectively leverage the useful prostate features. To verify the effectiveness of the 3D spatial and channel attention module, it is compared with the network without using the 3D spatial and channel attention module. The experimental results are shown in TABLE III. “3D EAGAN w/o 3D SCAM” represents the 3D spatial and channel attention module removed from the 3D EAGAN. “3D EAGAN w/ 3D SCAM” is kept in the 3D EAGAN. It can be observed that the use of the 3D spatial and channel attention module can slightly improve the used metrics.

\begin{table}[]
\caption{Evaluation of using the 3D spatial and channel attention module.}
\centering
\begin{tabular}{p{132pt}p{62pt}p{62pt}p{62pt}p{62pt}p{65pt}}
\hline
Method  & Dice(\%) & Jaccard(\%) & HD(mm) & Precision(\%) & Recall(\%)   \\ \hline
   3D EAGAN w/o 3D SCAM  & 92.07$\pm$0.91    & 85.04$\pm$1.04  & 5.20$\pm$0.92 & 92.04$\pm$1.26 & 91.75$\pm$ 1.76   \\
3D EAGAN w/ 3D SCAM	& \textbf{92.80$\pm$0.75}	& \textbf{87.01$\pm$0.42}	& \textbf{4.64$\pm$0.69}	& \textbf{93.11$\pm$0.62} &	\textbf{92.42$\pm$1.00} \\
        \hline
\end{tabular}
\end{table}

\subsubsection{Ablation study on the proposed edge enhance module}

The edge enhance module is utilized to guide the shallow layers of the edge-aware segmentation network to focus on the contour and edge information of prostates. To verify the effectiveness of the edge enhance module, ablation study experiments are performed. The experimental results are shown in TABLE IV. It can be observed that the use of the edge enhance module can improve the performance of the 3D EAGAN.

\begin{table}[]
\caption{Evaluation of using the edge enhance module.}
\centering
\begin{tabular}{p{115pt}p{65pt}p{65pt}p{65pt}p{65pt}p{65pt}}
\hline
Method  & Dice(\%) & Jaccard(\%) & HD(mm) & Precision(\%) & Recall(\%)   \\ \hline
   3D EAGAN w/o EEM  & 91.62$\pm$0.88    & 84.46$\pm$1.28  & 5.12$\pm$0.63 & 91.21$\pm$1.43 & 92.03$\pm$0.90   \\
3D EAGAN w/ EEM	& \textbf{92.80$\pm$0.75}	& \textbf{87.01$\pm$0.42}	& \textbf{4.64$\pm$0.69}	& \textbf{93.11$\pm$0.62} &	\textbf{92.42$\pm$1.00} \\
        \hline
\end{tabular}
\end{table}

To further verify the effectiveness of the edge enhance module, the visualization of feature maps extracted from the shallow layers of the edge-aware segmentation network is shown in FIG. 7. It can be observed that with the help of the edge enhance module, the prostate edge is more distinctive than the surrounding features in feature maps. Hence, shallow layers of the edge-aware segmentation network can pay attention to the edge of prostates.

\begin{figure}[htbp]
  \centering
  \includegraphics[width=6.5in]{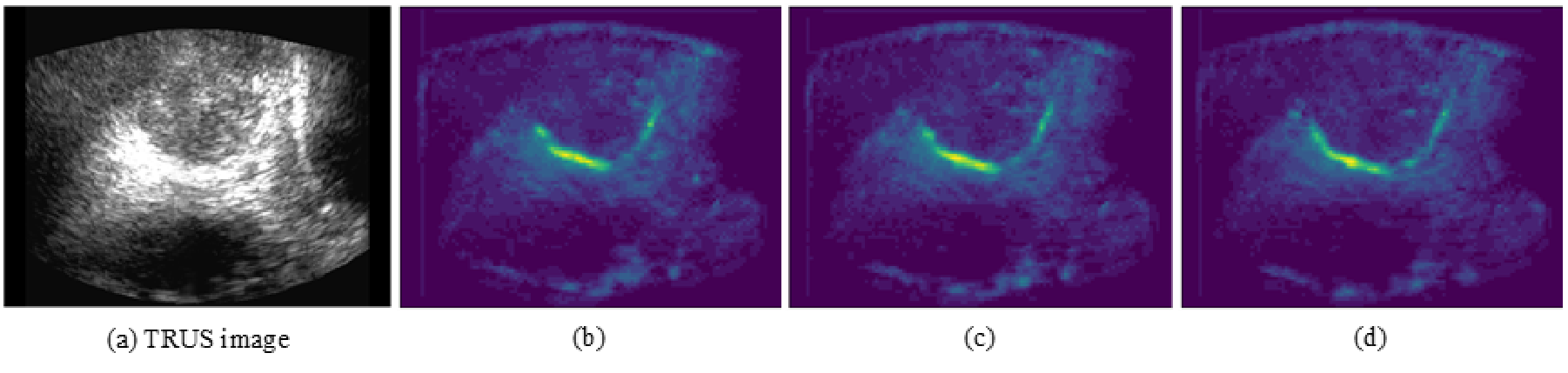}
  \caption{Visualization results of the feature maps extracted from the shallow layers of the edge-aware segmentation network. (a) is the input TRUS images; (b)-(d) are feature maps extracted from the shallow layers in the encoder module.}
\end{figure}

\subsubsection{Number of channel dimensions in the edge-aware segmentation network}

In the proposed 3D EAGAN, 3D convolutional layers are adopted to extract prostates in the 3D spatial domain. However, 3D convolutional layers inevitably increase computational complexity. To balance the segmentation performance and computational complexity, ablation study experiments on the number of channel dimensions in the edge-aware segmentation network are conducted. The experimental results are shown in TABLE V. Hence, the number of channel dimensions in the edge-aware segmentation network is limited to \{16, 32, 64, 128\} according to the experimental results.

\begin{table}[]
\caption{Evaluation of numbers of channel dimensions in the edge-aware segmentation network.}
\centering
\begin{tabular}{p{92pt}p{60pt}p{60pt}p{60pt}p{60pt}p{60pt}p{50pt}}
\hline
Channel dimension  & Dice(\%) & Jaccard(\%) & HD(mm) & Precision(\%) & Recall(\%) & Params(MB)  \\ \hline
   \{8, 16, 32, 64\}  & 90.48$\pm$0.89    & 84.33$\pm$0.79  & 8.11$\pm$0.83 & 89.45$\pm$0.69 & 88.34$\pm$ 1.28 & \textbf{73.92}   \\
\{16, 32, 64, 128\}	& \textbf{92.80$\pm$0.75}	& 87.01$\pm$0.42	& \textbf{4.64$\pm$0.69}	& \textbf{93.11$\pm$0.62} &	\textbf{92.42$\pm$1.00} & 75.10 \\
\{32, 64, 128, 256\} & 92.76$\pm$0.82 & \textbf{87.26$\pm$0.48} & 4.71$\pm$0.57 &93.21$\pm$0.66 & 92.40$\pm$0.87 & 78.77 \\
        \hline
\end{tabular}
\end{table}

\section{Conclusion}

In this paper, a 3D edge-aware attention generative adversarial network-based prostate segmentation method is proposed, which consists of an edge-aware segmentation network and a discriminator network. In the edge-aware segmentation network, the detail compensation module is proposed to introduce abundant detailed information to the network. In addition, an edge enhance module is proposed to guide shallow layers to pay attention to edge information prostates. Experimental results demonstrate the proposed method has achieved satisfactory results in 3D TRUS image segmentation of prostates.

\section{ACKNOWLEDGMENTS}
This work was partly supported by National Natural Science Foundation of China (Grant No.: 61601216), Science and Technology Key Research Project of Education Department of Jiangxi Province, China (Grant No.: GJJ2200114). The authors would like to thank Dr. Aaron Fenster and Lori Gardi from Western University, for their assistance with data collection and insightful comments.

\end{document}